# NEW STABLE METHOD TO SOLVE HEAT CONDUCTION PROBLEMS IN EXTREMELY LARGE SYSTEMS

ENDRE KOVÁCS[1]–ANDRÁS GILICZ

[1]University of Miskolc, Department of Physics,
3515 Miskolc-Egyetemváros
fizendre@uni-miskolc.hu

**Abstract:** We present a new explicit and stable numerical algorithm to solve the homogeneous heat equation. We illustrate the performance of the new method in the cases of two 2D systems with highly inhomogeneous random parameters. Spatial discretization of these problems results in huge and stiff ordinary differential equation systems, which can be solved by our novel method faster than by explicit or the commonly used implicit methods.

## 1. INTRODUCTION AND THE STUDIED PROBLEM

Experimental and numerical investigation of heat transfer in large-scale systems like chimneys of power plants [1], heat exchangers [2] and buildings [3] is a common problem for mechanical and heat engineers [4, 5]. In this paper we focus on only one of the mechanisms of heat transfer, the simplest Fourier-type heat conduction, but we hope that our results can be extended to heat transfer by radiation or even by convection.

Heat conduction phenomena are described by the heat equation, which is a second-order parabolic partial differential equation (PDE). Its homogeneous form is the following:

$$\frac{\partial T}{\partial t} = \alpha \Delta T,$$

where $\alpha = \frac{k}{c\rho}$ is the thermal diffusivity, k, c, and ρ is the heat conductivity, specific heat and (mass) density, respectively. The phrase 'homogeneous' does not mean that the medium is physically homogeneous (i.e. α is constant), but, instead, that we do not deal with heat source terms, so in this paper we examine only transient processes.

Most PDEs for real-life problems cannot be solved analytically. The process of numerical solution usually begins with the discretization of the space variables (there are few exceptions, for example Rothe's method). One has to divide the whole spatial domain into (small) blocks, during which (in case of the heat equation) one have to calculate two quantities for each block. The first one is the heat capacity *C* of the block:

$$C = c \cdot m = c\rho V \left[\frac{J}{K}\right],$$

where *m* is the mass, *V* is the volume of the block. Now one can obtain the (thermal) energy of a cell *j* as $C_j \cdot T_j$, where $T_j$ is the average temperature of the block. The second quantity is the heat/thermal conductance *U*, which can be approximated as

$$U_{ij} \approx k \frac{A_{ij}}{d_{ij}} \left[\frac{W}{K}\right],$$

where $A_{ij}$ is the surface area between the two blocks *i* and *j*, while $d_{ij}$ is the distance between the centres of the blocks. As we explained in our paper about analogies [6], quantities *C* and *U* are analogous to capacitance *C* and conductance (reciprocal resistance) *G = 1/R* in case of electrical RC circuits where – of course – the electric charge is the flowing quantity.

After spatial discretization according to the usual central formula for the second derivatives [7]

$$\frac{\partial^2}{\partial x^2} f(x_i, t_j) \approx \frac{f(x_{i+1}, t_j) - 2f(x_i, t_j) + f(x_{i-1}, t_j)}{(\Delta x)^2},$$

we obtain an ODE system which gives the time derivative of each temperature:

$$\frac{dT_i}{dt} = \sum_{j=neigh} \frac{U_{ij}}{C_i} (T_j - T_i),$$

where the summation is going over the neighbours of the block. In order to help the reader to visualize, we present the arrangement of the variables in *Figure 1* for a 2D system of 4 blocks.

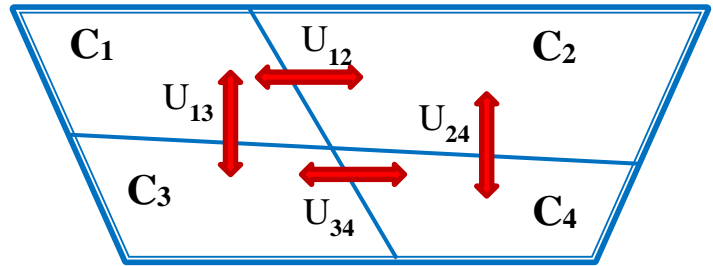

*Figure 1. Notations in the case of four blocks. The outer thicker line represents thermal isolation. We emphasize that the shape and arrangement of the blocks are not necessarily regular*

The ODE system in a matrix form for this small system:

$$\frac{d\vec{T}}{dt} = \begin{pmatrix} -\frac{U_{12}}{C_1} - \frac{U_{13}}{C_1} & \frac{U_{12}}{C_1} & \frac{U_{13}}{C_1} & 0 \\ \frac{U_{12}}{C_2} & -\frac{U_{13}}{C_2} - \frac{U_{24}}{C_2} & 0 & \frac{U_{24}}{C_2} \\ \frac{U_{13}}{C_3} & 0 & -\frac{U_{13}}{C_3} - \frac{U_{34}}{C_3} & \frac{U_{34}}{C_3} \\ 0 & \frac{U_{24}}{C_4} & \frac{U_{34}}{C_4} & -\frac{U_{24}}{C_4} - \frac{U_{34}}{C_4} \end{pmatrix} \vec{T}$$

One can see that the size of the matrix grows quadratically with the number of blocks, thus the number of elements of the matrix is inversely proportional with the 4$^{th}$ power of the diameter of the blocks (for a fixed system) and with the 6$^{th}$ power of the diameter in 3D. The absolute value of the matrix elements depends not only on the size (volume and surface) of the blocks but on the physical properties of the material like the specific heat and the thermal diffusivity as well. Since these parameters can largely vary from point to point, the magnitude of the matrix elements and therefore the eigenvalues can have a range of several orders of magnitude, which means it can be a severely stiff system.

Stiffness implies that conventional explicit methods are inappropriate because of unacceptably small timesteps. All available explicit integrators (with the possible exception of the Runge–Kutta–Chebyshev and the Alternating Direction methods) have a relatively small linear stability domain in the complex left half-plane [9, 10, 11]. This is the reason why they require unrealistically small step sizes for integrating stiff problems and they are rarely used in the industry. On the other hand, implicit methods require the solution of (usually nonlinear) algebraic equation systems at each time-step, moreover, it is not trivial to parallelize them. So when one has to quickly obtain an approximate result for a huge and stiff system, conventional methods provide no convenient solution. If the error tolerance is increased to enhance speed, explicit methods will diverge, while implicit ones still has to handle the huge matrices. Our task is to elaborate and new and easily parallelizable numerical algorithms and methods to solve these systems.

## 2. THE PROPOSED METHOD

We suggest the following simple formula to obtain the values of *T* at the end of the timestep using the values of *T* only at the beginning of the timestep:

$$T_i(t+h) = T_i(t) \cdot e^{-\frac{h}{\tau_i}} + \frac{\sum_{j=neigh} U_{ij} \cdot T_j(t)}{\sum_{j=neigh} U_{ij}} \cdot \left(1 - e^{-\frac{h}{\tau_i}}\right) \quad (1)$$

where $\tau_i = \dfrac{C_i}{\sum_{j=neigh} U_j}$ is the characteristic time of the block. With this formula we try to imitate the real processes in nature. In reality, if a system is thermally isolated, the temperature of each region of the system is approaching the equilibrium, which is the average temperature of the system. The speed of this process is proportional to the conductance between the region and the surroundings and inversely proportional to the heat capacity of the region. We tried to apply this physical principle, even if it implies that the method cannot be applied to other types of equations.

This method has the following advantages:
1) It is obviously explicit, one can calculate the new values without solving any kind of equation system or even without using matrices. It also implies that the process is easily parallelizable.
2) It is stable for heat conduction type problems, because the new value of the variable $T_i$ is the weighted average of $T_i$ and its neighbours $T_j$. Indeed, each coefficient in formula (1) is nonnegative and the sum of them is 1. Using this method one can be sure that the solution automatically follows the Maximum and Minimum principles [8], i.e. the extreme values of T occur among the initial values.
3) We state that it is convergent, i.e. if $h \to 0$ then the solution converges to the exact solution. At this moment this statement is based on numerical experiments, but we are working on the rigorous mathematical proof and it is planned to be published in a journal of applied mathematics.
4) It can be easily applied regardless of space dimension, lattice irregularity and inhomogeneity of the heat conduction medium.

We performed numerical tests on several systems, but here we present only 2 different examples.

### 3. THE FIRST EXAMPLE

The first system is a square-lattice, $N_x = 10$, $N_y = 10$. The capacities are $C_i = 10^{(2-5 \cdot rand)}$; while the conductances are $U_{xi} = 10^{(4 \cdot rand - 1)}, U_{yi} = 10^{(4 \cdot rand - 1)}$, where *rand* is a random number generated by the MATLAB uniformly in the (0, 1) interval for each block. It means that the capacities (the conductances) follow a log-uniform distribution between 0.001 and 100 (between 0.1 and 1,000). The initial temperatures followed a random function. The task is to solve this system for the temperatures between $t_0 = 0$s and $t_{FIN} = 1$s.

The stiffness ratio (the ratio of the [nonzero] eigenvalues of the matrix with the largest and smallest absolute value) is $6.9 \cdot 10^6$. For the explicit Euler method (which is equivalent to the forward-time central-space FTCS scheme), the maximum possible timestep is

$$h_{\text{MAX}}^{\text{E}} = \left| \frac{2}{\lambda_m} \right| = 3.85 \cdot 10^{-6} s ,$$

above this threshold instability necessarily occurs. Here $\lambda_m$ is (non-positive) eigenvalue of the matrix with the largest absolute-value.

In order to check our results, we used the ode45 Runge–Kutta–Dormand–Prince (RKDP) embedded adaptive-stepsize method which is built in MATLAB. First we chose strict error tolerance $('RelTol' = 10^{-7}, 'AbsTol' = 10^{-7})$, and with this, our PC needs 9.05 seconds to integrate the equation system. If we start to increase the error tolerance, the running time slowly decreases and can reach 8.37s (while the errors are increasing), but after a threshold, the program fails to converge. Implicit methods developed for stiff systems perform much better: ode23s and ode15s (which uses the Rosenbrock- and the BDF method, respectively, the letter *s* means that these codes were designed especially for stiff systems) solve the task in 0.5s quite precisely. However, as we mentioned before, increasing system size causes major problems for implicit methods, which will be illustrated in the second example.

Now let us try our method without parallelization. If we set the time-step size to h = 0.01, then our computer needs 0.0095s to solve it by our method. The result is presented in *Figure 2*. One can see that we managed to obtain a qualitatively good solution three orders of magnitude faster than the conventional explicit program. If we decrease h to 0.001, the orange line would be almost indistinguishable from the red one, while the running time would be still below 0.1s.

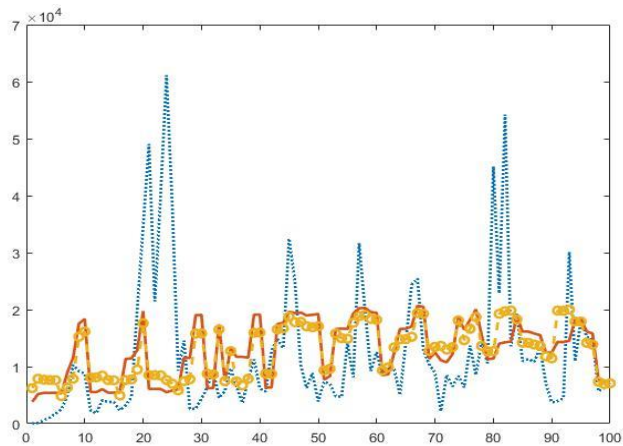

*Figure 2. The temperature as a function of the space variable. The blue dotted line represents the initial conditions, the red line is the high-precision solution while the orange circles are the values produced by our algorithm for h = 0.01*

A method is said to be *p*th order if the local error is $O(h^{p+1})$, or (equivalently for normal systems) if the global error is $O(h^p)$. On *Figure 3* we present 3 different kinds of the global error. The first one, MaxD is the maximum deviation from the exact (the high-precision) result. The second one, SumD is the sum of the deviations for all of the blocks. The 3$^{rd}$ one, EBE is the energy balance error. It can be calculated without the exact result due to the conservation of energy

$$EBE = \sum_{i=1}^{N} C_i \cdot \left(T_i(t = t_{FIN}) - T_i(t = t_0)\right)$$

One can see that the errors are decreasing slightly faster than the stepsize, thus it can be concluded that the convergence-rate of the method is (at least) one. The right side of the diagram also underpins the statement that the result is stable, as the error does not really increase for increasing stepsize.

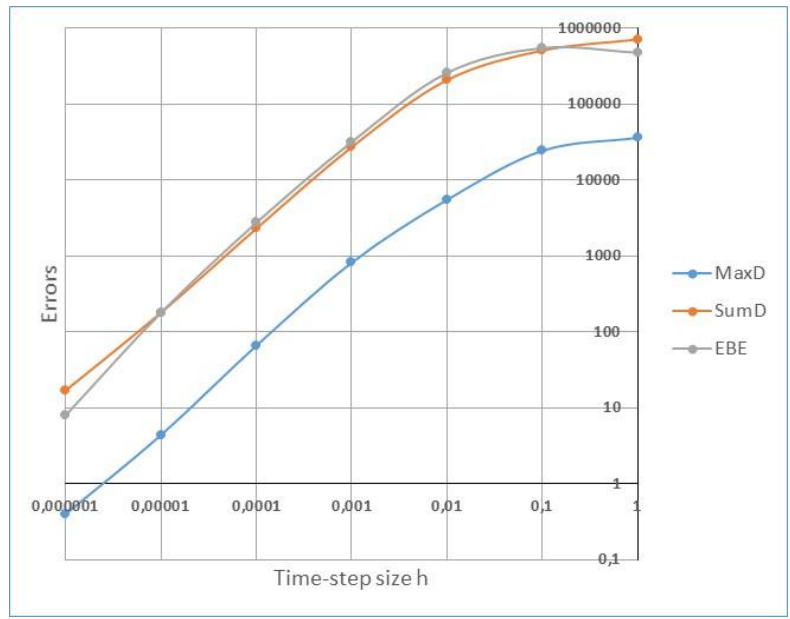

*Figure 3. Different kind of errors as the function of the timestep-size.*
*The blue line is the maximum difference, the orange is the sum of the differences*
*while the grey one is the absolute value of the energy balance error*

## 4. THE SECOND EXAMPLE

The second system is a rectangle-shaped lattice, $N_x = 400$, $N_y = 10$. The capacities are $C_i = 10^{(3-6 \cdot rand)}$; the distribution of the conductances $U_{xi} = 10^{(6 \cdot rand-2)}, U_{yi} = 10^{(6 \cdot rand-4)}$ is anisotropic. The initial temperatures follow a rectangular function:

$$T_i = \begin{cases} 100 & \text{if } 400 \leq i \leq 780 \\ 0 & \text{elsewhere} \end{cases}$$

The task is to solve this system for the temperatures between $t_0 = 0$s and $t_{FIN} = 100$s. The stiffness ratio of this problem is $1.36 \cdot 10^9$. For explicit Euler method, the maximum possible timestep is $h_{MAX}^E = 1.97 \cdot 10^{-7} s$. Thus explicit methods would require several hours or days to solve this problem, therefore we used implicit BDF method built in ode15s to provide us a reference solution. This MATLAB routine needs 712s to solve the problem with high precision. With loosening the error tolerance we could obtain results in 98s, but not sooner. On the other hand, our method needs roughly 0.0004s for one timestep, thus we can produce a rough but qualitatively good result in a few seconds, which means that we can beat the official routines if the main goal is not the precision but the speed. *Figure 4* reinforces our statement that the method is convergent with at least order 1.

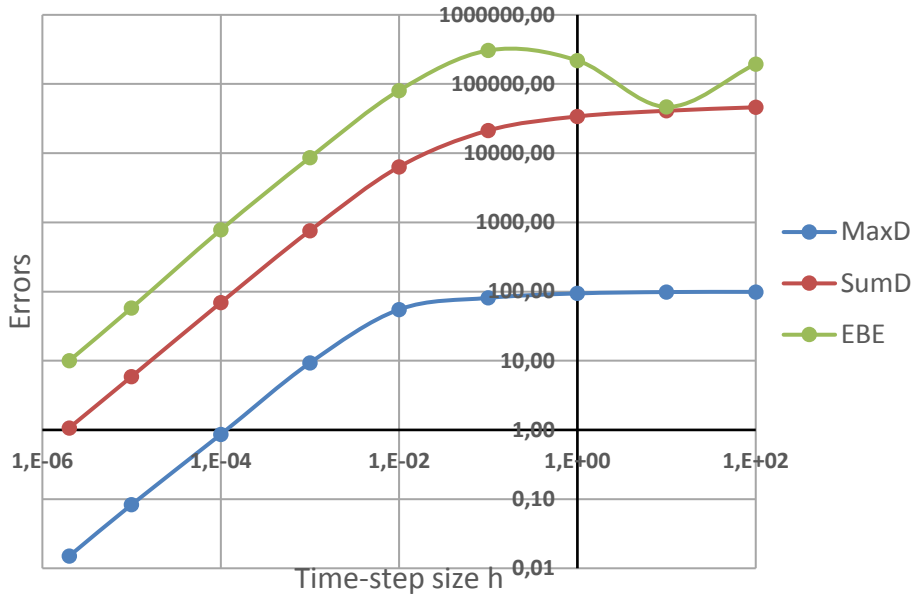

*Figure 4. Different kind of errors as the function of the timestep-size.*
*The blue line is the maximum difference, the orange is the sum of the differences*
*while the grey one is the absolute value of the energy balance error*

One can notice that the absolute value of the energy-balance error increases non-monotonously when the timestep is increased above $h = 1$. In fact, the EBE error changes sign in this region, which calls for caution about the usage of balance-errors as the main basis for error-estimation.

## 5. Summary

We presented a new numerical algorithm to solve the spatially discretized heat equation without external sources. This method is explicit, stable and convergent. We illustrated the performance of the method in case of two different systems with random parameters. The obtained data suggest that the larger the number of blocks (i.e. the variables in the ODE system) and the stiffness ratio is, the more significant the advantage of our method is, even without parallelization.

However, without fulfilling the following tasks, we could recommend this method only to solve special problems:
- Providing exact mathematical proof of the convergence.
- Working out how to handle source-terms, so that we can tackle problems other than transient problems.
- Elaborating modifications of the method for nonlinear versions of the heat equation, as the parameters like the specific heat usually depend on the temperature as well.
- Develop the adaptive stepsize control version of the method.
- Examine the possibilities for parallel programming of the method.

We are currently working on these projects and the proposed solutions will be published elsewhere.